\documentclass[titlepage]{tcibook}
\usepackage{fancyhea}
\usepackage{work}
\usepackage{bm}       
\usepackage{graphicx}
\usepackage{hyperref}      
\usepackage[usenames,dvipsnames]{color} 
\usepackage{amssymb}
\usepackage{amsmath}
\usepackage{chngcntr}
\usepackage{booktabs}
\usepackage{multirow} 
\usepackage{multicol}
\counterwithout{figure}{chapter}
\usepackage{titlesec}

\usepackage[utf8]{inputenc}

\newcommand{\nc}{\newcommand}  

\def\beq{\begin{equation}}
\def\eeq#1{\label{#1}\end{equation}}
\def\eeqn{\end{equation}}

\newenvironment{Eqnarray}%
   {\arraycolsep 0.14em\begin{eqnarray}}{\end{eqnarray}}
\def\beqa{\begin{Eqnarray}}
\def\eeqa#1{\label{#1}\end{Eqnarray}}
\def\eeqan{\end{Eqnarray}}

\nc{\ra}{\rightarrow}  
\nc{\slsh}{\slash\hspace*{-0.22cm}}
\def\Re{{\cal R \mskip-4mu \lower.1ex \hbox{\it e}\,}}
\def\Im{{\cal I \mskip-5mu \lower.1ex \hbox{\it m}\,}}

\nc{\vev}[1]{ \left\langle {#1} \right\rangle }
\nc{\bra}[1]{ \langle {#1} | }
\nc{\ket}[1]{ | {#1} \rangle }
\nc{\fb}{\,{\rm fb}^{-1}}
\nc{\ev}{{\rm eV}}
\nc{\kev}{{\rm keV}}
\nc{\Mev}{{\rm MeV}}
\nc{\gev}{{\rm GeV}}
\nc{\tev}{{\rm TeV}}
\nc{\mev}{{\rm MeV}}

\def\del{\partial}
\def\Dslash{\not{\hbox{\kern-4pt $D$}}}
\def\dslash{\not{\hbox{\kern-2pt $\del$}}}
\def\pslash{\not{\hbox{\kern-2pt $p$}}}
\def\ETmiss{ \not{\hbox{\kern-4pt $E$}}_T }

\def\msb{{\bar{\ssstyle M \kern -1pt S}}}

\definecolor{orange}{rgb}{1,0.3,0}

\def\lsim{\raise-.75ex\hbox{$\buildrel<\over\sim$}}
\newcommand{\RandD}{\ensuremath{\mathrm{R\& D\ }}}

\DeclareUrlCommand\email{\urlstyle{rm}}

\begin{document}

\def\bibname{References}

\bibliographystyle{utphys}  

\newenvironment{changemargin}[2]{%
\begin{list}{}{%
\setlength{\topsep}{0pt}%
\setlength{\leftmargin}{#1}%
\setlength{\rightmargin}{#2}%
\setlength{\listparindent}{\parindent}%
\setlength{\itemindent}{\parindent}%
\setlength{\parsep}{\parskip}%
}%
\item[]}{\end{list}}

\raggedbottom

\pagenumbering{roman}

\parindent=0pt
\parskip=8pt
\setlength{\evensidemargin}{0pt}
\setlength{\oddsidemargin}{0pt}
\setlength{\marginparsep}{0.0in}
\setlength{\marginparwidth}{0.0in}
\marginparpush=0pt


\renewcommand{\chapname}{chap:intro_}
\renewcommand{\chapterdir}{.}
\renewcommand{\arraystretch}{1.25}
\addtolength{\arraycolsep}{-3pt}

\pagenumbering{roman} 
\chapter*{Quantum Sensing \\ for \\ High Energy Physics }
\vskip -9.5pt
\hbox to\headwidth{%
       \leaders\hrule height1.5pt\hfil}
\vskip-6.5pt
\hbox to\headwidth{%
       \leaders\hrule height3.5pt\hfil}

\vskip 1.0cm 
  \begin{center}
   {\Large \bf
     Report of the first workshop to identify approaches and techniques in
      the domain of quantum sensing that can be utilized by
      future High Energy Physics applications to further the
      scientific goals of High Energy Physics.} \\
     \vskip 0.7in 
         {\large\bf 
           Organized by the Coordinating Panel for Advanced Detectors of
     the Division of Particles and Fields of the American Physical Society } \\ 
     \vskip 0.7in      
     {\Large\bf March 27, 2018 } 

\vskip 0.7in

%
%

\vskip 1.0in
       {\large
Karl van Bibber (UCB), 
Malcolm Boshier (LANL), 
Marcel Demarteau (ANL, co-chair)
Matt Dietrich (ANL), 
Maurice Garcia-Sciveres (LBNL)
Salman Habib (ANL), 
Hannes Hubmayr (NIST), 
Kent Irwin (Stanford), 
Akito Kusaka (LBNL), 
Joe Lykken (FNAL), 
Mike Norman (ANL), 
Raphael Pooser (ORNL), 
Sergio Rescia (BNL), 
Ian Shipsey (Oxford, co-chair), 
Chris Tully (Princeton).  

       }  
%
\end{center}
\eject

\setcounter{page}{1}

\begin{center}
  {\Large \bf Executive Summary}
\end{center}

The Coordinating Panel for Advanced Detectors (CPAD) of the APS Division
of Particles
and Fields organized a first workshop on Quantum Sensing for High Energy
Physics (HEP) in early December 2017 at Argonne National Laboratory.
Participants
from universities and national labs were drawn from the intersecting
fields of
Quantum Information Science (QIS), high energy physics, atomic, molecular and
optical physics, condensed matter physics, nuclear physics and materials
science. Quantum-enabled science and technology has seen rapid technical
advances and growing national interest and investments over the last few years.
The goal of the workshop was to bring the various communities together to
investigate pathways to integrate the expertise of these two disciplines
to accelerate the mutual advancement of scientific progress.
 
Quantum technologies manipulate individual quantum states and make use of
superposition, entanglement, squeezing and backaction evasion. Quantum
sensors exploit these quantum phenomena to make measurements
with a precision better than the Standard Quantum Limit, with the ultimate
goal of reaching the Heisenberg Limit.
New physics can be detected by causing tiny energy shifts in quantum
systems.  High energy physics experiments are only now beginning to leverage
quantum techniques that could enable significant improvements in
sensitivity. Resonance tools, for example, can powerfully probe for the new
particles predicted
by nearly all Beyond Standard Model theories that seek to explain some of
the biggest questions in particle physics today, such as the nature of dark
matter, dark energy, gravity and the hierarchy problem.
 
The workshop participants uniformly supported the assessment that great
opportunities exist for ambitious new initiatives.
The ascendancy of quantum tools for particle physics could be disruptive
for a certain class of studies,
and the participants embraced the discovery potential of these new
types of experiment. Most particle physicists who use accelerators, however, are not
well versed in the use of the tools and techniques from the quantum
sensing community. To maximize the impact of a new initiative in quantum
sensors for high energy physics (HEP) it is critical that the community begin
with as complete a knowledge of the current landscape as possible and a
recognition of how its execution may lead to very different projects
from the past. Establishing that knowledge could be
achieved with a targeted, in-depth survey to
map the current efforts in quantum sensing and extending it to quantum information science.
 
Given the inherent multi-disciplinary nature of QIS and
the potentially far-reaching implications of a Quantum
Sensor Initiative for high energy physics, structural changes could
be set in motion that need to be addressed. A new style of particle physics
experiment that is table-top in scale with multiple principal investigators
funded from a variety
of sources could tremendously accelerate scientific progress. However, this would
require a new level of coordination among funding agencies
within the Office of Science and NSF.
Furthermore, a coordinated effort needs to be put in place to help
high energy physicists, especially early career researchers,  become rapidly
fluent in the techniques of quantum information science.
 
Connecting HEP to QIS also promises to be a fruitful
line of research. One example is in the area of superconducting devices.
On the one hand, the facilities, expertise, and resources
of HEP superconducting detectors, for example, could provide a critical capability
for advancing QIS. On the other hand the development of new devices, materials 
and processes for 
QIS will inevitably lead to new techniques and devices e.g. for
low-noise cryogenic measurements of RF signals where unique QIS
capabilities such as squeezing and entanglement can enable
quantum and beyond quantum limited measurements that 
will extend the capabilities of detectors for high energy physics.
 
There is great promise in the use of quantum sensing for particle 
physics. We offer a set of suggestions to realize this promise in
section~\ref{chap:concl}.
The challenges faced by this emerging interdisciplinary
science may lead to a productive realignment and redefinition
of both scientific disciplines. We confidently predict that a targeted
initiative by DOE OHEP will act as a spur to enhance research that is
already beginning to
utilize quantum sensing in high energy physics 
and to discover other areas of particle physics where quantum
information and quantum sensing could have a powerful impact.


\eject

\clearpage
\newcounter{affilcount}

\newcommand{\affil}[1]{\refstepcounter{affilcount}\label{#1}}


\def\slactext{SLAC National Accelerator Laboratory, Menlo Park, California 94025, USA}
\affil{slac}

\def\anltext{Argonne National Laboratory, Lemont, Ilinois 60439, USA}
\affil{anl}

\def\fermilabtext{Fermi National Accelerator Laboratory, Batavia, Illinois 60510, USA}
\affil{fermilab}

\def\perimetertext{Perimeter Institute for Theoretical Physics, Waterloo, Ontario, N2L 2Y5 Canada}
\affil{perimeter}

\def\uchicagotext{University of Chicago, Chicago, Illinois 60637, USA}
\affil{uchicago}

\def\mittext{Massachusetts Institute of Technology, Cambridge, MA 02139, USA}
\affil{mit}

\def\uberkeleytext{University of California, Berkeley, Berkeley, CA 94720, USA}
\affil{uberkeley}

\def\jqitext{Joint Quantum Institute (JQI), University of Maryland, College Park, Maryland 20742, USA}
\affil{jqi}

\def\lanltext{Los Alamos National Laboratory, Santa Fe, New Mexico 87545, USA}
\affil{lanl}

\def\llnltext{Lawrence Livermore National Laboratory, Livermore, California 94550, USA}
\affil{llnl}

\def\harvardtext{Harvard University, Cambridge, Massachusetts 02138, USA}
\affil{harvard}

\def\pnnltext{Pacific Northwest National Laboratory, Richland, Washington 99354, USA}
\affil{pnnl}

\def\tumunichtext{Technical University of Munich, 80333 M¸nchen, Germany}
\affil{tumunich}

\def\lbnltext{Lawrence Berkeley National Laboratory, Berkeley, California 94720, USA}
\affil{lbnl}

\def\northwesterntext{Northwestern University, Evanston, Illinois 60208 USA}
\affil{northwestern}

\def\torontotext{University of Toronto, Toronto, Ontario M5S, Canada}
\affil{toronto}

\def\iheptext{Institute of High Energy Physics, Shijingshan Qu, Beijing Shi, 100049, China}
\affil{ihep}

\def\stanfordtext{Stanford University Physics Department, Stanford, California 94305, USA}
\affil{stanford}

\def\nisttext{National Institute of Standards \& Technology, Gaithersburg, Maryland, 20899 USA}
\affil{nist}

\def\niutext{Northern Illinois University, DeKalb, Illinois 60115 USA}
\affil{niu}

\def\csueastbaytext{California State University, East Bay, Hayward, California 94542, USA}
\affil{csueastbay}

\def\tokyotext{University of Tokyo, Bunkyo, Tokyo 113-8654, Japan}
\affil{tokyo}

\def\ornltext{Oak Ridge National Laboratory, Oak Ridge, Tennessee 37830 USA}
\affil{ornl}

\def\jilatext{JILA, University of Colorado, Boulder, Colorado 80309, USA}
\affil{jila}

\def\browntext{Brown University, Providence, Rhode Island 02912, USA}
\affil{brown}

\def\bnltext{Brookhaven National  Laboratory, Upton, New York 11973, USA}
\affil{bnl}

\def\jpltext{Jet Propulsion Laboratory, Pasadena, California 91109, USA}
\affil{jpl}

\def\brandeistext{Brandeis University, Waltham, Massachusetts 02453, USA}
\affil{brandeis}

\def\minnesotatext{The University of Minnesota, Minneapolis, Minnesota55455, USA}
\affil{minnesota}

\def\caltechtext{California Institute of Technology, Pasadena, California 91125, USA}
\affil{caltech}

\def\oxfordtext{University of Oxford, Oxford OX1 2JD, United Kingdom}
\affil{oxford}

\def\princetontext{Princeton University, Princeton, NJ 08544, USA}
\affil{princeton}

\def\iittext{Illinois Institute of Technology, Chicago, Illinois 60616, USA}
\affil{iit}


\def\INFNPDtext{Istituto Nazionale di Fisica Nucleare, Sezione di Padova,  35131 Padova , Italy}
\def\INFNCAtext{Istituto Nazionale di Fisica Nucleare, Sezione di Cagliari, 09042 Cagliari, Italy}
\def\INFNGEtext{Istituto Nazionale di Fisica Nucleare, Sezione di Genova, 16146 Genova, Italy}
\def\INFNLNFtext{Istituto Nazionale di Fisica Nucleare, Laboratori Nazionali di Frascati, 00044 Frascati, Italy}
\def\INFNCTtext{Istituto Nazionale di Fisica Nucleare, Sezione di Catania, 92125 Catania, Italy}
\def\INFNRMtext{Istituto Nazionale di Fisica Nucleare, Sezione di Roma, 00185 Roma, Italy}
\def\INFNLEtext{Istituto Nazionale di Fisica Nucleare, Sezione di Lecce, 73047, Italy}
\def\INFNRMtretext{Istituto Nazionale di Fisica Nucleare, Sezione di Roma Tre, 00146 Roma, Italy}
\def\INFNLNStext{Istituto Nazionale di Fisica Nucleare, Laboratori Nazionali del Sud, 92125 Catania , Italy}
\def\UNISStext{Universit\`a di Sassari, 07100 Sassari, Italy}
\def\UNIRMtext{Universit\`a di Roma La Sapienza, 00185 Roma, Italy}
\def\UNIMEtext{Universit\`a di Messina, 98122 Messina, Italy}
\def\UNIGEtext{Universit\`a di Genova, 16100 Genova, Italy}
\def\UNISAtext{Universit\`a del Salento, 73100 Lecce, Italy}
\def\UNIPDtext{Universit\`a di Padova, 35122 Padova, Italy}
\def\UNIRMTVtext{Universit\`a di Roma Tor Vergata, 00173 Roma, Italy}
\def\INFNRMtvtext{Istituto Nazionale di Fisica Nucleare, Sezione di Roma Tor Vergata, 00173 Roma, Italy}
\def\UNIMAINZtext{JGU Mainz, Institute for Nuclear Physics and PRISMA Cluster of Excellence, 55128 Mainz, Germany}
\def\INFNTOtext{Istituto Nazionale di Fisica Nucleare, Sezione di Torino, 10125 Torino, Italy}
\def\orsaytext{Institut de Physique Nucl\'eaire, CNRS-IN2P3, Univ. Paris-Sud, Universit\'e Paris-Saclay, 91406 Orsay Cedex, France}
\def\ethZtext{Institute for Particle Physics, ETH Zurich, 8093 Zurich, Switzerland}
\def\MITtext{Massachusetts Institute of Technology, Cambridge, MA 02139, USA}
\def\CERNthtext{CERN Theory Department, CH-1211 Geneva 23, Switzerland}
\def\yangstonybrooktext{C.N.~Yang Institute for Theoretical Physics, Stony Brook University, Stony Brook, NY 11794, USA}
\def\stonybrooktext{Stony Brook University, Stony Brook, NY 11794, USA}
\def\riversidetext{Department of Physics and Astronomy, University of  California, Riverside, California 92521, USA}
\def\kentuckytext{Department of Physics and Astronomy, University of Kentucky, Lexington, Kentucky 40506, USA}
\def\calstatetext{California State University, Los Angeles, Los Angeles, California 90032, USA}
\def\yaletext{Yale University, New Haven, CT 06520, USA}
\def\ODUtext{Old Dominion University,  Norfolk, VA 23529, USA}
\def\cornelltext{Cornell University,  Ithaca, NY 14853, USA}
\def\LOSALAMOStext{Los Alamos National  Laboratory, Los Alamos, NM 87545, USA}
\def\UDAVIStext{University of California, Davis, Davis, CA 95616, USA}
\def\WMtext{College of William and Mary, Williamsburg, VA 23185, USA}
\def\princetontext{Princeton University, Princeton, NJ 08544, USA}
\def\uvictoriatext{University of Victoria, Victoria, BC V8P 5C2, Canada}
\def\TRIUMFtext{TRIUMF,  Vancouver, BC  V6T 2A3, Canada}
\def\dubnatext{Joint Institute for Nuclear Research, 141980 Dubna, Russia}
\def\UCALSBtext{University of California, Santa Barbara, Santa Barbara, CA 93106, USA}
\def\tuebingentext{Institute of Theoretical Physics, University of Tuebingen, D-72076, Tuebingen, Germany}
\def\tomsktext{Department of Physics, Tomsk State University, 634050 Tomsk, Russia}
\def\tomskBtext{Laboratory of Particle Physics, Mathematical Physics Department, Tomsk Polytechnic University, 634050 Tomsk, Russia}

\begin{center}
  {\Large \bf Authors}
\end{center}
\bigskip

\begin{raggedright}


Zeeshan Ahmed\textsuperscript{\ref{slac}},
Yuri Alexeev\textsuperscript{\ref{anl}},
Giorgio Apollinari\textsuperscript{\ref{fermilab}},
Asimina Arvanitaki\textsuperscript{\ref{perimeter}},
David Awschalom\textsuperscript{\ref{uchicago}},
Karl K.~Berggren\textsuperscript{\ref{mit}},
Karl Van Bibber\textsuperscript{\ref{uberkeley}},
Przemyslaw Bienias\textsuperscript{\ref{jqi}},
Geoffrey Bodwin\textsuperscript{\ref{anl}},
Malcolm Boshier\textsuperscript{\ref{lanl}},
Daniel Bowring\textsuperscript{\ref{fermilab}},
Davide Braga\textsuperscript{\ref{fermilab}},
Karen Byrum\textsuperscript{\ref{anl}},
Gustavo Cancelo\textsuperscript{\ref{fermilab}},
Gianpaolo Carosi\textsuperscript{\ref{llnl}},
Tom Cecil\textsuperscript{\ref{anl}},
Clarence Chang\textsuperscript{\ref{anl},\ref{uchicago}},
Mattia Checchin\textsuperscript{\ref{fermilab}},
Sergei Chekanov\textsuperscript{\ref{anl}},
Aaron Chou\textsuperscript{\ref{fermilab}},
Aashish Clerk\textsuperscript{\ref{uchicago}},
Ian Cloet\textsuperscript{\ref{anl}},
Michael Crisler\textsuperscript{\ref{fermilab}},
Marcel Demarteau\textsuperscript{\ref{anl}},
Ranjan Dharmapalan\textsuperscript{\ref{anl}},
Matthew Dietrich\textsuperscript{\ref{anl}},
Junjia Ding\textsuperscript{\ref{anl}},
Zelimir Djurcic\textsuperscript{\ref{anl}},
John Doyle\textsuperscript{\ref{harvard}},
James Fast\textsuperscript{\ref{pnnl}},
Michael Fazio\textsuperscript{\ref{slac}},
Peter Fierlinger\textsuperscript{\ref{tumunich}},
Hal Finkel\textsuperscript{\ref{anl}},
Patrick Fox\textsuperscript{\ref{fermilab}},
Gerald Gabrielse\textsuperscript{\ref{harvard},\ref{northwestern}},
Andrei Gaponenko\textsuperscript{\ref{fermilab}},
Maurice Garcia-Sciveres\textsuperscript{\ref{lbnl}},
Andrew Geraci\textsuperscript{\ref{northwestern}},
Jeffrey Guest\textsuperscript{\ref{anl}},
Supratik Guha\textsuperscript{\ref{anl}},
Salman Habib\textsuperscript{\ref{anl}},
Ron Harnik\textsuperscript{\ref{fermilab}},
Amr Helmy\textsuperscript{\ref{toronto}},
Yuekun Heng\textsuperscript{\ref{ihep}},
Jason Henning\textsuperscript{\ref{anl}},
Joseph Heremans\textsuperscript{\ref{anl}},
Phay Ho\textsuperscript{\ref{anl}},
Jason Hogan\textsuperscript{\ref{stanford}},
Johannes Hubmayr\textsuperscript{\ref{nist}},
David Hume\textsuperscript{\ref{nist}},
Kent Irwin\textsuperscript{\ref{slac},\ref{stanford},\ref{slac}},
Cynthia Jenks\textsuperscript{\ref{anl}},
Nick Karonis\textsuperscript{\ref{niu},\ref{anl}},
Raj Kettimuthu\textsuperscript{\ref{anl}},
Derek Kimball\textsuperscript{\ref{csueastbay}},
Jonathan King\textsuperscript{\ref{uberkeley}},
Eve Kovacs\textsuperscript{\ref{anl}},
Richard Kriske\textsuperscript{\ref{mit}},
Donna Kubik\textsuperscript{\ref{fermilab}},
Akito Kusaka\textsuperscript{\ref{lbnl},\ref{tokyo}},
Benjamin Lawrie\textsuperscript{\ref{ornl}},
Konrad Lehnert\textsuperscript{\ref{jila}},
Paul Lett\textsuperscript{\ref{nist}},
Jonathan Lewis\textsuperscript{\ref{fermilab}},
Pavel Lougovski\textsuperscript{\ref{ornl}},
Larry Lurio\textsuperscript{\ref{niu}},
Xuedan Ma\textsuperscript{\ref{anl}},
Edward May\textsuperscript{\ref{anl}},
Petra Merkel\textsuperscript{\ref{fermilab}},
Jessica Metcalfe\textsuperscript{\ref{anl}},
Antonino Miceli\textsuperscript{\ref{anl}},
Misun Min\textsuperscript{\ref{anl}},
Sandeep Miryala\textsuperscript{\ref{fermilab}},
John Mitchell\textsuperscript{\ref{anl}},
Vesna Mitrovic\textsuperscript{\ref{brown}},
Holger Mueller\textsuperscript{\ref{uberkeley}},
Sae Woo Nam\textsuperscript{\ref{nist}},
Hogan Nguyen\textsuperscript{\ref{fermilab}},
Howard Nicholson\textsuperscript{\ref{anl}},
Andrei Nomerotski\textsuperscript{\ref{bnl}},
Mike Norman\textsuperscript{\ref{anl}},
Kevin O'Brien\textsuperscript{\ref{uberkeley}},
Roger O'Brient\textsuperscript{\ref{jpl}},
Umeshkumar Patel\textsuperscript{\ref{anl}},
Bjoern Penning\textsuperscript{\ref{brandeis}},
Sergey Perverzev\textsuperscript{\ref{llnl}},
Nicholas Peters\textsuperscript{\ref{ornl}},
Raphael Pooser\textsuperscript{\ref{ornl}},
Chrystian Posada\textsuperscript{\ref{anl}},
Jimmy Proudfoot\textsuperscript{\ref{anl}},
Tenzin Rabga\textsuperscript{\ref{anl}},
Tijana Rajh\textsuperscript{\ref{anl}},
Sergio Rescia\textsuperscript{\ref{bnl}},
Alexander Romanenko\textsuperscript{\ref{fermilab}},
Roger Rusack\textsuperscript{\ref{minnesota}},
Monika Schleier-Smith\textsuperscript{\ref{stanford}},
Keith Schwab\textsuperscript{\ref{caltech}},
Julie Segal\textsuperscript{\ref{slac}},
Ian Shipsey\textsuperscript{\ref{oxford}},
Erik Shirokoff\textsuperscript{\ref{uchicago}},
Andrew Sonnenschein\textsuperscript{\ref{fermilab}},
Valerie Taylor\textsuperscript{\ref{anl}},
Robert Tschirhart\textsuperscript{\ref{fermilab}},
Chris Tully\textsuperscript{\ref{princeton}}, 
David Underwood\textsuperscript{\ref{anl}},
Vladan Vuletic\textsuperscript{\ref{mit}},
Robert Wagner\textsuperscript{\ref{anl}},
Gensheng Wang\textsuperscript{\ref{anl}},
Harry Weerts\textsuperscript{\ref{anl}},
Nathan Woollett\textsuperscript{\ref{llnl}},
Junqi Xie\textsuperscript{\ref{anl}},
Volodymyr Yefremenko\textsuperscript{\ref{anl}},
John Zasadzinski\textsuperscript{\ref{iit}},
Jinlong Zhang\textsuperscript{\ref{anl}},
Xufeng Zhang\textsuperscript{\ref{anl}},
Vishnu Zutshi\textsuperscript{\ref{niu}},



\normalsize

\parskip=4pt
\begin{center}
\noindent
\textsuperscript{\ref{slac}}\slactext \\
\noindent
\textsuperscript{\ref{anl}}\anltext \\ 
\noindent
\textsuperscript{\ref{fermilab}}\fermilabtext \\
\noindent
\textsuperscript{\ref{perimeter}}\perimetertext \\
\noindent
\textsuperscript{\ref{uchicago}}\uchicagotext \\
\noindent
\textsuperscript{\ref{mit}}\mittext \\
\noindent
\textsuperscript{\ref{uberkeley}}\uberkeleytext \\
\noindent
\textsuperscript{\ref{jqi}}\jqitext  \\
\noindent
\textsuperscript{\ref{lanl}}\lanltext  \\
\noindent
\textsuperscript{\ref{llnl}}\llnltext  \\
\noindent
\textsuperscript{\ref{harvard}}\harvardtext  \\
\noindent
\textsuperscript{\ref{pnnl}}\pnnltext \\
\noindent
\textsuperscript{\ref{tumunich}}\tumunichtext  \\
\noindent
\textsuperscript{\ref{lbnl}}\lbnltext \\
\noindent
\textsuperscript{\ref{northwestern}}\northwesterntext \\
\noindent
\textsuperscript{\ref{toronto}}\torontotext \\
\noindent
\textsuperscript{\ref{ihep}}\iheptext \\
\noindent
\textsuperscript{\ref{stanford}}\stanfordtext \\
\noindent
\textsuperscript{\ref{nist}}\nisttext  \\
\noindent
\textsuperscript{\ref{niu}}\niutext \\
\noindent
\textsuperscript{\ref{csueastbay}}\csueastbaytext \\
\noindent
\textsuperscript{\ref{tokyo}}\tokyotext  \\
\noindent
\textsuperscript{\ref{ornl}}\ornltext  \\
\noindent
\textsuperscript{\ref{jila}}\jilatext  \\
\noindent
\textsuperscript{\ref{brown}}\browntext \\
\noindent
\textsuperscript{\ref{bnl}}\bnltext \\
\noindent
\textsuperscript{\ref{jpl}}\jpltext \\
\noindent 
\textsuperscript{\ref{brandeis}}\brandeistext  \\
\noindent
\textsuperscript{\ref{minnesota}}\minnesotatext \\
\noindent
\textsuperscript{\ref{caltech}}\caltechtext  \\
\noindent
\textsuperscript{\ref{oxford}}\oxfordtext  \\
\noindent
\textsuperscript{\ref{princeton}}\princetontext  \\
\noindent
\textsuperscript{\ref{iit}}\iittext \\

\end{center}
\normalsize


\parskip=8pt

\end{raggedright}

\clearpage

\begin{center}
  {\Large \bf Preface}
\end{center}
\bigskip

This report is the product of the first workshop on Quantum Sensing for High Energy Physics,
held at Argonne National Laboratory, December 12 -- 14, 2017. Quantum Information Science
(QIS) and its instrumentation constitute a rapidly-developing interdisciplinary field,
whose novel approaches to fundamental science, enabled by understanding and manipulation 
of the uniquely quantum phenomena of superposition, entanglement, and squeezing could
hold great promise to answer the fundamental questions posed by HEP. 
Sizeable investments are being made to advance QIS and significant progress has been
achieved in developing the technologies and techniques for QIS.
The workshop sought to identify approaches and techniques in the domain of quantum
sensing that offer opportunities to devise powerful probes to explore physics at new
energy scales relevant to the mission of particle physics. It is noted that the work in
QIS specifically as it relates to quantum computing was not part of the workshop. 
 
The techniques used in QIS span a broad spectrum, many of
which are not familiar to the HEP community. This workshop reached out to other
communities and organizations inviting scientists with the relevant expertise to
help inform and guide the HEP community in utilizing quantum technology
and presenting other non-accelerator opportunities for doing particle particle physics.
The workshop started
with a genuinely pedagogical review of the state of the art of principal techniques
used in the field today and the underlying theory. This was followed by talks by
practitioners using these principal techniques and their variants. Several
panel discussions were held. These examined the role of the national laboratories
and the future actions  the field of HEP might consider to leverage the synergies
with QIS. Using a web-based document management tool, 
participants shared their ideas and presenters were asked to submit a
one-page summary of their presentation. This report summarizes the key
findings of the workshop and suggests next actions that could be taken by
the field of High Energy Physics.
The individual summaries of the presenters are available on the agenda server
for the meeting~\cite{qs_indico}.

\tableofcontents


\def\as#1{[{\bf AS:} {\it #1}] }


\eject
\pagenumbering{arabic} 
\setcounter{page}{1}
\chapter{Introduction}
\label{chap:intro}
\bigskip

Particle physics explores the fundamental constituents of energy and matter and the nature of space and time.
It applies that knowledge to understand the birth, evolution and fate of the Universe.
Through careful measurement, observation and deduction the field has developed a remarkably successful
prevailing theory - the Standard Model of Particle Physics (SM) that is highly predictive and has been
rigorously tested, in some cases to one part in ten billion.  

However, what we know and what we are made of, described by the SM is just the tip of the iceberg.
The sense of mystery has never been greater in particle physics.  At the heart of the SM is a Higgs
field and particle that is an enigma. The Higgs is the first elementary spin~0 particle found in nature.
It has a mass unprotected from quantum corrections that will act to increase it by a factor of about 17~orders
of magnitude unless there is a remarkable degree of fine tuning between the corrections to about one part
in $10^{34}$, the Hierarchy problem. Particle physicists study the Higgs in detail and use it as
a tool for discovery at the Large Hadron Collider (LHC); this work is in its infancy.

Astrophysical data has demonstrated the existence of dark matter and dark energy, and shown
that the primordial balance between matter and antimatter was erased as the universe evolved.
Dark Matter holds our universe together but it challenges our understanding.
For every gram of ordinary matter in the universe the other four are dark.
We don't know the dark quantum.
There is no particle in the SM that has the correct properties to be the Dark Matter
particle, the Dark Matter problem. Numerous approaches are being pursued to detect Dark Matter
directly, and these are complemented by searches at the LHC, surveys of large-scale structure in
the Universe and attempts to observe high-energy particles from Dark Matter annihilation or
decay in or around our galaxy. So far, a solution to the Dark Matter problem has not been found.

Dark Energy drives our universe apart and accounts for 70\% of the mass-energy of the universe.
It is a mystery. A term arising in the theory of general relativity, known as the cosmological
constant, $\Lambda$, has the properties of Dark Energy; however, connecting it to the fundamental
properties of quantum fields is problematic. The ever-present ``vacuum energy" of the quantum fields
in the Standard Model may be connected to $\Lambda$, but would naively predict a value of $\Lambda$
about 120~orders of magnitude larger than is observed.  While potential solutions to the hierarchy
problem are numerous, addressing the cosmological constant problem is proving extremely challenging.
Experimental probes of Dark Energy include optical and radio surveys of large scale structure in the
cosmos and measurements of the cosmic microwave background (CMB). Furthermore, precision clock
measurements seek to see direct effects of Dark Energy in the laboratory.

How did matter survive the Big Bang? Some phenomena must have produced a small asymmetry between matter
and antimatter (the CP-violation problem) so that when every billion matter particles annihilated a
billion anti-matter particles, a single matter particle remained.
CP-violation is  probed through table-top EDM searches, experiments at dedicated b-factories and the LHC,
in kaon experiments, in accelerator-based neutrino experiments, and in the search for neutrinoless
double beta decay. 

What powered Cosmic Inflation, the exponential expansion of the early universe by a factor of at least $10^{78}$ in volume, starting about $10^{-36}$~s after the big bang until about
$10^{-33} -10^{-32}$~s? It may be due to a scalar field - particle physics at the
Planck scale. Following the inflationary period, the universe continued to expand at a
slower rate, until Dark Energy became important.
CMB surveys look for the signature of inflation. These surveys have grown in scope from
satellites like WMAP to Planck and, in the future, the proposed LiteBIRD and from
ground-based surveys that will find their most mature expression in the proposed
CMB Stage IV mission. 

There are many more questions, the solutions to which may in some cases be related to the mysteries  described above.  They include why are there so many types of particle? Why do the particles have such a large range of masses? Why does the pattern of particles (generations) repeat three times?
Why do neutrinos, that are massless in the SM, have mass at all? 

A wide range of theoretical solutions has been proposed to address the mysteries described in the previous paragraphs including Supersymmetry, theories of extra spatial dimensions, and theories that unify the strong force with the electroweak  force.  These theories invariably predict the existence of new particles close to the Higgs mass. This fact is one of the driving forces for the  LHC program. New, light particles that couple only very weakly to ordinary matter including axions, axion-like particles, dark photons, dilatons, moduli, milli-charged particles, and other exotic particles, also arise in extensions of the SM and are not excluded by current experimental data.
Many of the particles predicted in extensions of the SM are candidates for Dark Matter
and range in mass from more than 100~GeV (e.g. WIMPS)
to $10^{-22}$~eV. It is noted that light Dark Matter is best considered as a wave
instead of a particle since the particle number density is extremely high.
Extensions of the SM also provide additional sources of CP violation. The effects of the predicted new particles may also be seen in precision measurements at lower energy and in rates of rare processes. For example they may produce non-zero nucleon and electron electric dipole moments (EDMs), the violation of charged lepton flavor symmetry
allowing, e.g. a muon, to transform into an electron, which is forbidden in the SM,
and deviations from the SM predictions for the muon and electron anomalous
gyromagnetic ratios.  

Searches for these new particles and their associated quantum fields directly, or
indirectly  through their effects on SM particles, probe fundamental physics at
energies above the TeV scale directly explored by the LHC,  and in some cases
probe the Planck scale ($(10^{19}$~GeV).
To date, no new particles beyond those in the SM have been detected
directly. 
Given the cost and complexity of large-scale experiments, there is growing interest
in complementary approaches to address the fundamental questions, which lie at the
heart of modern particle physics and cosmology. 

Remarkably, it has become possible to meaningfully explore the mysteries of particle
physics with experiments small enough to fit in a laboratory room due to the
development of ultra-precise quantum sensors. Some examples of experiments where
quantum sensors can make a decisive contribution include searches for EDMs,
Dark Matter, violations of fundamental symmetries, and the signature of
Cosmic Inflation, and studies of Dark Energy, and gravity. Current EDM experiments
exploit methods closely related to those developed for quantum sensor technologies.
Anticipated near-future developments in quantum techniques are opening the prospect
for further, rapid improvement.
The development of networks of quantum sensors will be optimized to search, e.g. for
classes of non-WIMP Dark Matter (which in some cases can also be identified with the
field responsible for Dark Energy); violations of Lorentz and CPT symmetry;
and temporal or anomalous spatial variations or oscillations of fundamental physical
constants or interactions.
Atom interferometry has demonstrated space-time curvature across a single particle’s
wave function, allows a sensitive test of the equivalence principle, provides an
alternative way to search for light Dark Matter and could provide midband
(10~mHz-10~Hz) atomic gravitational wave detection bridging the sensitivity
gap between LIGO and the proposed LISA satellite.
Quantum sensors also probe quantum mechanics itself. For example, long-distance
breakdowns in standard quantum mechanics interpretations of entanglement
and other phenomena can be explored.
 
These new small-scale experiments rely on approaches fundamentally different compared
to traditional large particle physics experiments. Breakthrough ideas, coupled with
rapid advances in atomic, molecular, and optical physics techniques and
quantum-limited measurement devices, are leading to a surge in progress.
For example, some quantum sensors rely on quantum-mechanical resonance behavior,
which enables exquisitely sensitive measurement of tiny energy shifts, shifts
that can be caused by the existence of new quantum fields with exactly the
properties needed to solve several of the mysteries of particle physics.
Through recent progress in quantum technologies, it is now possible to resolve the zero-point motion of high-quality oscillators, and consequently, to search for these fields at a rate limited by fundamental quantum noise~\cite{Brubaker:2017}. Furthermore, it is not only possible, but in some cases essential, to circumvent this quantum noise. For the case of the QCD axion for example, searches already operate at the quantum limit using Josephson Parametric Amplifiers (JPAs) and they will need to circumvent this limit by using quantum squeezing and microwave photon-counting 
techniques~\cite{Zheng:2016}.

The small scale experiments proposed to investigate new fields and particles predicted
in beyond SM theories rely heavily on current and near-future advances in the field
of quantum sensors.  Sensors can consist of many different signal transduction
mechanisms (such as the displacement of a micromechanical device, or the change in
the index of refraction of a medium, etc.). The signal can be read off from the
transducer using various physical readout fields, which may be quantized
(for example, atomic or optical fields used to read out the phase shift in an
interferometer). Signal fields of interest to HEP consist of magnetic, electric,
gravimetric, and others.

The fundamental measurements associated with many of these detection modalities
have undergone extensive optimization to minimize classical noise sources and
reach the Standard Quantum Limit. In many sensors, including atomic or optical
interferometers, magnetometers, and accelerometers, quantum noise reduction, or
``squeezing'' can be used to reduce noise in these devices to a point below the
Standard Quantum Limit, to a lower bound constrained only by general minimum
uncertainty states which saturate the Heisenberg Uncertainty Principle. Other
types of sensors including nanoscale devices such as nitrogen vacancy centers
in diamond, or superconducting circuits can be used to detect intense fields
at the nanoscale while coupling with quantum readouts. A reduction in variance
of 100 times (20~dB) has been observed in atom fields, while optical fields have
demonstrated a similar 15~dB reduction.

For sensors which exhibit high losses in their readout fields (which reduces squeezing),
other quantum effects, such as entanglement, quantum illumination,
and measurement induced nonlinearity via post-selection can be used to boost
dynamic range or reduce the noise floor as well.
An ideal scenario for application of these states is the regime in which back-action,
which results in an increase in noise due to the interaction of the readout field
with the signal transducer, is prevalent. In sensors that operate at the Standard
Quantum Limit, and in which back action is not a dominant source of noise, the
signal to noise is dependent on the intensity of the readout field (the optical
power in the case of light, or the number of atoms in an interferometer in the
case of atomic sensors, for instance). However, in the back-action limited regime,
signal to noise cannot be further improved by increasing the intensity of the
readout field, because back action noise would commensurately increase.
Quantum states, such as those that exhibit squeezing, are required to reduce
the noise floor further.  A reduction in the noise floor of a few decibels
could reduce the integration time of a search.

It is important to recognize that quantum information and sensing science do not
represent a panacea for every sector of high energy physics, but
it has the potential to have a fairly broad impact. Today light Dark Matter
searches are already profiting from the quantum paradigm. Moving forward we
expect further sectors of particle physics to benefit, but to achieve this
will take significant effort. New approaches need to be developed for a
successful marriage of quantum information science and high energy physics.
It will require studies to frame the fundamental particle physics questions
using quantum sensing techniques. New technologies might have to be grafted,
optimally correlated states will have to be defined, and how squeezing,
superposition or entanglement can be incorporated in the experiments to
reach the Heisenberg limit demand serious investments both in intellectual
as well as capital resources bringing the right communities together.

The remainder of the report is organized as follows:  section~2 describes the
principles of quantum sensing and quantum sensing techniques and technologies;
section~3 describes the
potential role of the National Laboratories and the program of each relevant
to quantum sensing; section~4 concludes the report including some suggestions
for the Office of High Energy Physics to consider.     
 
\chapter{Experimental Techniques}
\bigskip

Quantum sensors have the potential to significantly impact particle physics.
Investments in quantum techniques are necessary for high-energy and table-top
precision physics experiments
to reach their scientific potential. In this section we describe overall principles
of quantum sensing, and describe specific opportunities in several different quantum
sensor technologies and experimental techniques.

\section{Principles of quantum sensing}\label{sec:Principles}

\subsection{Exploiting quantum correlations} \label{sec:QuantumCorrelations}
The Heisenberg Uncertainty Principle limits the sensitivity of some measurements used
in high-energy physics, including field measurement, position sensing, magnetometry,
and interferometry. The limit placed on simultaneous measurements of two non-commuting
quantities (such as the amplitude and phase, or the cosine and sine quadrature of an
electromagnetic signal) is referred to as the Standard Quantum Limit (SQL).
Quantum Sensors can exploit quantum correlations to make measurements better than the SQL,
improving the science reach of HEP experiments.
Measurement protocols can be used to measure below the SQL by taking advantage of
\emph{squeezing}, \emph{entanglement}, \emph{backaction evasion}, \emph{photon counting},
and other techniques.

At higher photon energy, photon counting has been done for many years, but photon counting
at low photon energy (e.g. microwaves) is still a very new technique that has promise for
HEP experiments. Photon counting can also be considered a quantum protocol. A photon
counter measures the number of quanta, but not any information about phase. Photon
counting is thus not limited by the SQL. If the source is in a Fock state
(an eigenstate of photon number), the measurement can in principle be done repeatedly
(referred to as quantum non-demolition) without introducing shot noise. 

All of these quantum protocols are deeply interrelated. Different protocols are used
under different circumstances, but all use the resource of quantum correlations to
make measurements below the SQL. High-energy physics experiments are only now beginning
to leverage the quantum techniques that will enable improvements in sensitivity beyond
this limit. Quantum correlations enable many different sensor technologies, including
those that directly measure signals with sensitivity below the SQL, and those that
exploit quantum correlations to achieve higher sensitivity in interferometry.
Ultimately, quantum correlations allow interferometry to improve inversely as the
number of photons (the Heisenberg Limit) rather than as the square-root of the number
of photons (the SQL).

\subsection{Quantum sensor systems} \label{sec:QuantumSystems}
The full impact of quantum sensors on HEP science will only be reached as full
quantum systems are implemented. Some quantum systems require modifications to
operation not just at the level of individual sensors, but also in their integration,
readout, networking, and data acquisition.

A single quantum sensor can only take advantage of quantum correlations over the extent
of its spatial probability distribution, while a physical field of interest may be
spread over a wider area.
To measure a larger area will require quantum networks, which take advantage of 
correlations across an array of sensors,
linking them to each other with quantum mechanical means, such as optical entanglement.
For example, atomic clocks can be networked together to form an even more precise
network of clocks.
It is also likely that future quantum sensor networks will be comprised of multiple,
disparate small-scale quantum devices. This can be achieved by linking individual
quantum sensors using photons as a quantum interconnect to allow quantum information
exchange between subsets of sensing devices. Popular matter qubit
systems, often used as quantum sensors, such as trapped ions, neutral atoms,
quantum dots, or single
dopants in silicon can be readily coupled to optical photons using optical
non-linearities in narrow frequency
bins~\cite{lukens_frequency-encoded_2017, lu_electro-optic_2018}.
Photonic protocols offer
linear scaling of optical resources with off-the-shelf telecom components and enable
highly parallel entanglement of multiple frequency modes within a single optical
fiber. Continuous-variable quantum networks can also be deterministically generated
with non-linear optics media, and can link and entangle the quadratures of remote
supeconducting resonators.

Quantum techniques can also be used to improve the performance of sensor arrays already
used in HEP experiments with a Gaussian (thermal) sensor output. For example,
the exploitation of squeezing and entanglement in the amplifier chain can be used to
implement larger arrays of transition-edge sensors (TES)~\cite{Irwin1995} similar to
those used in searches for dark-matter candidates (e.g. low-mass WIMP searches
following on SuperCDMS) and in searches for the signature of inflation in the
CMB (e.g. CMB-S4).
The exploitation of quantum correlations in networks of TES devices can improve
the dynamic range of the readout elements, increasing multiplexibility, leading
to larger, more sensitive arrays with lower power dissipation, as will be described
in section~\ref{sec:TES}.

\section{Quantum techniques and technologies} \label{sec:QuantumTechniques}
\subsection{Detection of electromagnetic fields} \label{sec:EMfields}

The sensitive measurement of electromagnetic fields plays an important role in
many high-energy physics experiments. One of the earliest applications of quantum
sensors to HEP is in the measurement of the coupling of ultralight dark matter
(including axions and hidden photons) to electromagnetism.
In these experiments, the signal results from the coupling of two harmonic oscillators,
i.e. between the dark matter field (with Compton frequency $f_C \approx m c^2$
determined by the rest mass $m$ of the dark matter) and a tuned electromagnetic
resonator (in the case of the axion, in the presence of a dc magnetic field).

Coupling to electromagnetic resonators is used to search for dark matter over a large frequency range, 100~${\rm Hz} < f < 20$~GHz.
In the microwave range ($f > 300$~MHz) resonant cavities are cooled to the ground state, and squeezing or photon counting with qubits can search with sensitivity better than the SQL. At lower frequencies ($f < 300$~MHz), lumped-element LC resonators are used. While these resonators are in a thermal state, the search bandwidth is greatly increased by
the use of quantum correlations, including backaction evasion, to measure with
sensitivity below the SQL.

Four technologies are under development as quantum sensors for electromagnetic axion and hidden-photon detection. Quantum squeezing, which is ready to accelerate QCD axion searches today, will be the first method used to circumvent quantum noise. A squeezed-state receiver uses two JPAs, one to prepare a squeezed state and one to amplify that state after it has interacted with the axion detection apparatus. The quantum-noise limited HAYSTAC (Haloscope At Yale Sensitive To Axionic Cold Dark Matter) experiment~\cite{haystac}
will deploy a squeezed state receiver to speed up the search starting in early 2018.

Superconducting transmon qubits are under development as microwave photon counters
to accelerate microwave resonant searches. Although there are some substantial
technical hurdles to overcome, the fact that photon counting is much less sensitive
to loss than squeezing gives it a long-term technical advantage.
Photon counting techniques can accelerate searches by orders of magnitude for
frequencies above 10~GHz and if the apparatus operates below 100~mK. 

Microwave Superconducting Quantum Interference Device (SQUID) frequency upconverters
are under development for quantum sensing
at frequencies below 300 MHz in the Dark Matter Radio experiment~\cite{dmradio}.
Photon counting is not useful in this frequency range, as the resonator is in a thermal state. However, quantum correlations can still be exploited by techniques including backaction evasion to measure below the SQL, greatly increasing sensitive bandwidth, and accelerating dark matter searches down to 100 Hz.

Detectors of single optical photons are useful for probes of coupling to dark-matter
candidates at higher masses. Superconducting transition-edge sensors (TES) and
superconducting-nanowire single-photon detectors (SNSPD) are broadband photodetectors
with relatively large area, low noise, high efficiency, and low energy
threshold~\cite{Berggren1}. These capabilities are instrumental for detection of
low-mass dark matter particles, whose absorption or scattering in a bulk volume
primarily produces just a single photon of \`a priori unknown energy~\cite{Berggren2}.
The device efficiency can be $>$ 90\%, and the detector can be sensitive from the UV
through the infrared band. Detection of low-probability events require detectors
with extraordinarily low dark counts---below a few per day. 

These detectors also have promise in experiments that rely on the quantum nature
of light to improve sensitivity in precision measurements, for example, experiments
that might use entangled states of light to improve precision in interferometry.

In addition, the recent discovery of materials characterized by a relativistic
dispersion (Dirac semimetals) offers the possibility of new materials
for dark matter detection with sensitivities that are predicted to exceed those
using more traditional materials~\cite{hoch}.
In particular, these materials can be tuned to open up a small band gap,
suppressing thermal noise.  And their relativistic dispersion leads to a
dielectric response that differs significantly from that of metals and
semiconductors.  The prediction is that for dark matter particles with a
mass in the range of a keV to a MeV, such materials should be ideal detectors,
even exceeding what can be done with superconductors, as well as semiconductors
such as germanium and silicon that have much larger energy gaps.

Trapped ions can also act as field sensors. 
Trapped, laser-cooled ions, with a high degree of control over both the internal
and motional states, are capable of sensing extremely weak foces ($< 1$~yN)
and electric fields ($< 1$~nV/m)~\cite{maiwald,hempel,gilmore}.
Electric field sensing below $\approx$1~nV/m enables searches for hidden-photon
dark matter in the 50~kHz to 5~MHz frequency interval corresponding to the
range of motional frequencies for trapped ions.

\subsection{Nuclear Magnetic Resonance} \label{sec:Spin}

One of the principal ways in which dark matter composed of bosonic fields interacts
with standard model particles is through interactions with intrinsic
spins~\cite{Kimball4, Kimball5, Kimball6}. The techniques of magnetic resonance
are ideally suited to detect such effects, since magnetic-field interactions have
characteristics similar to dark-matter bosonic-field interactions.
Nuclear Magnetic Resonance (NMR) techniques are used
in the Cosmic Axion Spin Precession Experiment (CASPEr)
Wind experiment for detecting spin precession caused by bosonic
fields~\cite{Kimball7}.
The coupling of the bosonic dark-matter field to nuclear spins has the form of a
pseudo-magnetic field oscillating at the boson’s Compton frequency. If the external
bias magnetic field is set to a value such that the nuclear spin splitting matches
this frequency, a resonance condition is achieved, and the nuclear spins tilt and
undergo Larmor precession. A sensitive magnetometer, such as a SQUID, placed next
to the sample, detects the oscillating transverse magnetization.
The polarized nuclear spin samples used as dark matter detectors in CASPEr rely on
spin coherences, and quantum measurement protocols and new quantum sensor technology
are being developed to optimize sensitivity and bandwidth.

\subsection{Quantum systems using transition-edge sensors} \label{sec:TES}
Superconducting transition-edge sensors (TESs) are used in HEP experiments to search for WIMPs (e.g. SuperCDMS) and to search for the signature of inflation, search for light relics, and constrain neutrino mass with the CMB. At first look, it is not obvious that quantum squeezing and entanglement can improve the performance of TES detectors, but as mentioned in
section~\ref{sec:QuantumSystems}, quantum squeezing and entanglement can be used to improve the performance of TES arrays, even though the output of a TES is a thermal distribution. 

TES detectors are presently used in HEP experiments either in non-multiplexed implementations, or multiplexed with time-division and frequency-division techniques. All HEP experiments using TES detectors presently use SQUID amplifiers. These readout technologies are mature and have been deployed on multiple HEP experiments (CDMS, SuperCDMS, CMB, etc.). The microwave SQUID multiplexer \cite{Irwin2004} is an emerging technology that expands the multiplexing factor by an order of magnitude. Microwave SQUID multiplexers are the baseline readout technique for the Simons Observatory, and are under consideration for CMB-S4 and future light-dark-matter searches.

In microwave SQUIDs, the signal from a TES is encoded only in the phase of a ``flux-ramp modulation'' signal in a microwave resonator. Present microwave SQUID TES readout electronics measure both the phase and amplitude of the signal in the microwave resonator, so they are limited by the SQL. However, if techniques including squeezing or two-resonator entanglement are implemented, the phase of the signal can be measured below the SQL. This reduces the flux noise of the TES measurement, which increases its dynamic range, reducing the bandwidth required by each pixel, and reducing the power required in the resonator. Consequently, it would sometimes be possible to make larger TES arrays with lower power
using quantum sensor techniques. 

For example, some future experiments to measure light WIMPs require very large arrays of detectors with small absorber volumes. If squeezing is used to increase the dynamic range of the microwave SQUID, reducing the bandwidth of each signal, then more pixels can be instrumented on each output coaxial cable. Thus, quantum techniques have the potential to enable larger instrumented absorber volumes in low-mass WIMP experiments.

\subsection{AMO techniques} \label{sec:AMO}
Experiments with ultra-cold atoms and ions benefit from extraordinarily precise
control of both the
atomic system and its environment. Frequency ratio measurements of optical
atomic clocks exhibit systematic uncertainties and measurement stabilities
extending below $1 \times 10^{-17}$. At the same time, these measurements have provided the most
stringent constraints on several extensions to the standard model, including possible time variation of the fundamental constants~\cite{Hume2,Hume3}.
Optical atomic clocks may soon be employed as sensors to search for dark
matter~\cite{Schleier7,Schleier8} or detect gravitational waves~\cite{Schleier9}.
Figures of merit for such sensors include their speed, bandwidth, spatial resolution, and dynamic range.  Quantum entanglement can offer dramatic enhancements in any one (or more) of these figures of merit, by maximizing the precision attainable in fixed time with a finite number of particles~\cite{Schleier10}.

In the context of quantum information processing (QIP), maximally
entangled states of N ions have been generated for N as high as 14~\cite{Hume5}.
Future measurements involving optical lattice clocks and trapped ion clocks
could operate close to
the Heisenberg limit, realizing a state-of-the-art measurement with the maximal
quantum enhancement.
Atomic ensembles have also been used extensively for magnetometry, which can play
a vital role in EDM searches and axion detectors. Squeezed light has been shown
to enable even further improved sensitivity in these atomic magnetometers~\cite{otter,wolfgram}.  

Advances in control and characterization of quantum many-body systems, driven in part
by goals in en\-tangle\-ment-enhanced sensing, have also opened new opportunities
for quantum simulations relevant to high-energy
physics~\cite{Schleier16, Schleier17, Schleier18}.
A particularly promising direction is to experimentally investigate the connections
between entanglement and space-time geometry conjectured under gauge/gravity
duality~\cite{Schleier19,Schleier20}.  

\subsection{Helium} \label{sec:Helium}
The properties of cryogenic helium isotopes also lend themselves to quantum sensing.
Since the axion couples to nuclear spin, which is proportional to the nuclear magnetic
moment, the axion coupling can be treated as a fictitious magnetic field.
The Axion Resonant InterAction Detection Experiment (ARIADNE) aims to detect
axion-mediated spin-dependent interactions between an unpolarized source mass and a
spin-polarized $^3$He
low-temperature gas~\cite{Geraci5,Geraci6}. This is accomplished by spinning an
unpolarized tungsten mass sprocket near a $^3$He vessel. As the teeth of the sprocket
pass by the sample at the nuclear Larmor precession frequency, the magnetization
in the longitudinally polarized He gas begins to precess about the axis of an
applied field. This precessing transverse magnetization is detected with a SQUID.
The $^3$He sample acts as an amplifier to transduce the small fictitious
magnetic field into a larger real magnetic field detectable by the SQUID,
or a more advanced quantum sensor. The ultimate limit is set by quantum spin
projection noise in the sample itself~\cite{Geraci5,Geraci6}.

It was also noted a number of years ago that a quantum condensate
can be used to amplify weak signals to macroscopic proportions,
specifically exploiting the unique properties of superfluid $^3$He to detect the
electric dipole moment~\cite{leggett}.

\subsection{EDM measurements} \label{sec:EDM}
Although no EDM of an elementary particle has been observed yet, that property can be measured with exquisite
precision. Modern day EDM measurements probe certain physics at energy scales
competitive or even exceeding those available at the LHC~\cite{Dietrich1,Dietrich2}.
Experiments in this area are only now beginning to fully leverage the quantum
techniques that will enable huge improvements in the next generation of experiments.
Methods like stimulated Raman adiabatic passage have been recently implemented in
the thorium monoxide and radium experiments, and promise one to two decades of
improved sensitivity.
Cold-atom based EDM experiments, such as the radium experiment, could someday
also benefit from spin squeezing, which would reduce detection noise below the
shot-noise limit, thereby providing an additional one to two orders of magnitude
in sensitivity.
Rapid advances are being made in techniques to cool and trap the heavy polar
molecules that are exquisitely sensitive probes of the electron EDM.
The availability of trapped molecules or molecular fountains will extend
interaction times by orders of magnitude, which will in turn increase experimental
sensitivity to the eEDM by orders of magnitude.
 
Neutron EDM experiments are based on either cold neutron beams or trapped ultra-cold (few mK) neutrons, produced at research reactors or spallation neutron sources. Using Ramsey’s technique of separated oscillatory fields, neutron spins are exposed to simultaneous magnetic and electric fields. An EDM would cause a shift in Larmor-frequency depending on the relative alignment of the applied fields. After about 60 years of continuous improvements, the experimental sensitivity for the neutron’s EDM now correspond to an energy resolution of $10^{-22}$ eV. Already this sensitivity corresponds to an energy scale of up to 100 TeV, depending on the new physics model. 

\subsection{Atom interferometers} \label{sec:AtomI}
Atom interferometers are just beginning to reach their potential as quantum sensors
for high-energy physics and gravitational waves.
For example, the MAGIS (Matter wave Atomic Gradiometer Interferometric Sensor)
detector concept is based on a new kind of atomic sensor that is a hybrid between
an atomic clock and an atom interferometer.  Dilute clouds of ultracold atoms at
both ends of a long baseline act as inertial test masses and local clocks,
and laser light propagates between the atoms. The lasers are used to implement
light pulse atom interferometry locally at each end of the baseline, and the resulting
interference patterns are compared.  This differential signal is sensitive to
variations in both the light travel time and the atomic energy level spacings,
while the influence of laser frequency noise is suppressed as a common mode.
The light flight time is recorded by the accumulation of phase by these atoms,
which also serve as precise differential clocks.
MAGIS-100 is sensitive to time-variations in the atomic energy levels caused
by couplings to ultralight dark matter, since such energy level shifts change
the phase accumulation by the separated atomic clocks.
Ultralight dark matter candidates with mass in the range of $10^{-13}$~eV to $10^{-16}$~eV
can cause time-varying atomic energy levels in the 0.1~Hz to 10~Hz frequency range.  

Atom interferometers are also useful in measurements of the fine-structure constant,
$\alpha$, another precision probe of fundamental physics~\cite{g-2a,g-2b}. 
Knowing $\alpha$ allows predicting the gyromagnetic anomaly of the electron
and is a particularly broad search for new physics. Already now, it sets
stringent limits on several proposed particles in a region of mass and coupling
constant where accelerator experiments are currently insensitive.
The measurement is based on the recoil velocity of an atom that has been
``kicked” by a photon of known momentum.
They have been used to perform the most accurate measurement of $\alpha$ to date at
0.20~parts per billion (ppb)~\cite{mueller}, and are targeting an
accuracy of 0.005 - 0.02 parts per billion (ppb). 
With experiments of this nature, new physics can be probed broadly and deeply
and be sensitive to a number of new particles in unprobed regions of mass
and coupling constant.

\chapter{The Role of National Laboratories}
\label{chap:labs}


\bigskip

Currently, universities and several national laboratories and institutes
play the leading role in the advancement of QIS
and the development of its technologies.
They also continue to play the dominant role in particle physics using quantum sensing.
Only recently is high energy physics exploring the use of QIS techniques and
technologies into its experimental program.
Universities 
are the training ground for the next generation of physicists that will apply, adopt, and
enhance these technologies to address fundamental particle physics questions.
The role of the national laboratories in this area is, however, expected to increase.
Although many experiments are currently laboratory-scale ``tabletop'' experiments,
the rapid advances in the field will naturally lead to larger
systems where the inherent expertise of the national laboratories may be essential 
to advance the field. This section summarizes our view of
several national laboratories on their potential contribution to this science. 

The workshop provided a forum for the DOE laboratories to share their capabilities and
interests in quantum sensors and testbed operations. Participating laboratories were
Argonne National Laboratory (ANL), Brookhaven National Laboratory (BNL),
Fermi National Accelerator Laboratory (FNAL),
Los Alamos National Laboratory (LANL), Lawrence Berkeley National Laboratory (LBNL),
Lawrence Livermore National Laboratory (LLNL), Oak Ridge National Laboratory (ORNL) and 
Stanford Linear Accelerator National Laboratory (SLAC). One DOC laboratory, the
National Institute of
Standards and Technology (NIST), also participated in the forum. It is evident that 
the national laboratories can play a unique role in the development of quantum sensors
due to their deep expertise and technical capabilities. Furthermore, the laboratories
have the capability for scale-up of systems that is absent from most universities. 

The picture that emerged from the discussions is that the National Laboratories are
an ideal platform for bringing QIS to bear on future
HEP programs, given their multi-disciplinary nature.  All of the participating DOE
laboratories commented on the fact that they not only have HEP programs, but QIS
programs as well.  The synergy of the combined National Laboratory efforts are 
expected to lead to the next generation of detectors and sensors, and will allow
experiments like dark matter searches to be performed that are presently not possible
because current technology has
reached its sensitivity limits.  The labs have the ability to perform the
necessary \RandD work and have experience in scaling up to the arrays
needed for doing large scale experiments.  For scale up, many of them are also
able to exploit nanofabrication and synthesis facilities,
including several who have BES Nanoscale Science Research Centers
(ANL, BNL, LANL, LBNL, ORNL), as well as similar facilities connected with partner
universities, with Berkeley, Chicago, Stanford and Stony Brook being notable examples.
The national labs are also adept at large project management, a key skill given the
challenges in scaling up quantum technologies, as well as the necessary system integration.

The multi-disciplinary labs can also exploit synergy with other co-located facilities
in need of similar technologies for next generation detector arrays.
These include the BES light sources (ALS, APS, LCLS, NSLS-II, SSRL),
BES neutron sources (HFIR, SNS), and various nuclear physics facilities (ATLAS, CEBAF, RHIC).
A number of experiments being planned by the Office of Nuclear Physics (EDM and NLDBD,
for instance) have similar technological requirements to those of planned HEP experiments.
The capabilities of NIST are well known, having been exploited in the past for HEP
projects connected with the cosmological microwave background and dark matter searches,
and will be equally important moving forward given their overall expertise in quantum
technologies.

Several labs have mentioned their plans in quantum communications, including quantum
networks and the quantum repeaters associated with them (ANL, BNL, FNAL, ORNL), and
their goal of leveraging these efforts to benefit their HEP programs.  A number of them
also have co-located high performance computing facilities (ANL, LBNL, LLNL, ORNL, PNNL),
which could be exploited not only for quantum simulation emulators but also for 
computational chemistry to address materials science challenges. 
In addition, the
new ASCR Quantum Testbeds program (currently deployed at LBNL and ORNL) provides access
to quantum algorithms and computing that could make fundamental progress in numerical
solutions of QCD and other hard HEP problems that have been resistant to solution.
In connection with this, most of the labs with co-located HEP programs also have
programs in qubit development. 
The co-location of QIS and HEP leads to the realistic possibility of quantum sensor designs
being tailored to match needs and maximize impact on selected future HEP programs.

In the remainder of this section each laboratory has summarized their capabilities and future plans.
Many of the future activities center around the desire for quantum sensing
to achieve low noise detection at the single photon limit with fast time capabilities
that could bring CMB and various dark matter and neutrino searches to the next level
of sensitivity.  ORNL is currently exploiting squeezed states and quantum entanglement
to improve sensitivity beyond the Standard Quantum Limit in sensing platforms that can
be adapted for HEP needs.  Others work on
superconducting RF cavities with high quality factors, which, when coupled to qubits,
can lead to new capabilities.  FNAL is also looking into cold atom interferometry and
LANL is investigating atomtronic circuits as new tools that could be exploited in
collaboration with NIST's extensive experience in cold atom technologies.

\section{ANL}
The field of quantum sensing is wide in scope, extending across the scales, and perspectives
of various fields, from the nanosciences to cosmology.  Industrial R\&D seems 
ill-equipped for research of this magnitude, and while universities have excellent subject
matter expertise, they are not the ideal places for integrating such activity over large
and multidisciplinary projects. The government laboratories, on the other hand, are ideal
for this type of research. They possess the large scale research facilities, the
government-laboratory university consortia, the user facilities that act as waterholes,
and the extensive synthesis and massively parallel characterization capabilities that
push the limits of instrumentation.

At Argonne these include the Advanced Photon Source's ability to perform dynamical
strain mapping (in three dimensions) of active quantum structures such as qubits.
The laboratory has extensive know how and nanofabricational skills in superconducting
technology,
elaborate synthesis facilities and theory and modeling for materials discovery for
new qubits. The lab is one of the founders of the Chicago Quantum Exchange (involving ANL,
FNAL and the University of Chicago),
and the
Quantum Factory-extensive synthesis capabilities is sited in the Materials Science Division
and the Institute for Molecular Engineering.  Addtionally, the Center for
Nanoscale Materials,
a DOE user facility, is ramping up its efforts in unique user facilities for
quantum materials,
while the High Energy Physics (HEP) and Physics (PHY) divisions are beginning to evaluate
quantum sensing and detection approaches for their specific problems.
In particular, HEP has an extensive program in developing TES arrays for CMB
studies (SPT-3G), which is now being mirrored by PHY for a program in
neutrinoless double beta decay (CUPID).  PHY also has a program in atom
traps (radium EDM project).

\section{BNL}
The BNL Instrumentation Division has thirty years of experience in developing low noise,
low temperature electronics systems for HEP, including ASICs. It also has modern
facilities for silicon sensor R\&D, nanofabrication and metrology.
BNL is hosting the US ATLAS experiment with its Tier~I computing and Phase~II
Upgrade activities, and plays important roles in neutrino experiments, LSST and
Belle~II. Outside of HEP, BNL is home to Relativistic Heavy Ion Collider and its
experiments, NSLS-II and Computing Science Initiative providing expertise and
infrastructure, which in many respects is congruent with the HEP frontiers. 

Cross fertilization between HEP and QIS is going both ways and there are good
examples of HEP products used for QIS. At BNL, this is an intensified, time stamping
camera with single photon sensitivity and nanosecond timing resolution.
The design of the silicon sensor in the camera, in particular its thin entrance
window, was inspired by the fully depleted astronomical CCDs used in LSST,
while the readout chip is a product of the Medipix collaboration at CERN employing
technologies developed for the LHC experiments. 

The camera is used in collaboration with the Stony Brook University quantum science
group for characterization of sources of single photons and quantum memories,
and for measurements of temporal and spatial correlations of entangled photons
in quantum networks. The camera also allows investigations on photons simultaneously
entangled in polarization and spatial OAM (orbital angular momentum) modes.
Combination of these hyper-dense encoded states with room temperature quantum
memories seems feasible and would be an exciting development for quantum cryptography.
There is also interest to use the camera for imaging of single photons in studies of
non-equilibrium quantum systems and other types of quantum imaging. The advances in
the sensor technology, such as LGAD and SPAD sensors with internal amplification,
are on the radar of the LHC upgrades and could be used to produce new imaging
instruments with single photon sensitivity for the QIS applications.

A variety of new single-photon, low temperature detector technologies with
infrared sensitivity, excellent timing resolution and a high signal-to-noise
ratio are being developed and evaluated in QIS community. A good example of a
QIS technology, which may find applications in HEP are the superconductive
nanowire single photon detector (SNSPD). Many scintillators emit in the
blue-UV spectrum while the noble liquid emission wavelength is even shorter.
As was reported at the workshop SNSPD have the potential of QE close to 100\%
and should be capable to operate at very short wavelength. For HEP applications,
an array of SNSPDs could be built on top of an array of comparators with micro-bump
interconnects to register the liquid Argon scintillation light.
A suitable high temperature superconductor, compatible with the LAr operating
temperature would be especially attractive.

There are also opportunities for theoretical HEP investigations of how quantum
simulations and computations can be used for Quantum Field Theory (QFT) including
its non-Abelian and Beyond the Standard Model variations.

\section{FNAL}

Fermilab has been involved in superconducting detector development for many years.
With sensor development targeting the sub-eV energy scale of axion dark matter,
neutrino masses, dark energy, and low mass WIMP recoils, the program also aims to
improve the sensitivity of experimental cosmology by integrating superconducting
sensors into highly multiplexed arrays with applications in precision CMB measurements,
high statistics galaxy redshift surveys, and lower mass axion dark matter searches.
In each of these science targets, the expected low signal power requires sub-Kelvin
cryogenic operation of detectors in order to reduce thermal background noise to
acceptable levels.  In this regime, the small binding energy of Cooper pairs in
low-T$_c$ superconductors provides an attractively small energy quantum for precision
bolometry and micro-calorimetry with low energy threshold. Moreover, the quantum
mechanics of individual Cooper pairs enables quantum non-demolition measurements of
individual microwave photons to bypass the Standard Quantum Limit on low-noise
amplifiers and potentially achieve orders of magnitude in noise reduction in these
sensors.  The availability of low temperature, high field test stands also enables
the detailed study of superconducting materials and surface treatment methods to
understand the fundamental microphysics of energy loss mechanisms which limit the
performance of superconducting detectors as well as that of superconducting RF
accelerator cavities.

Fermilab operates a suite of cryogenic detector and materials test facilities,
which currently include adiabatic demagnetization refrigerators, helium-3 cryo-coolers,
a 4K, 9T Quantum Designs PPMS system, and several dilution refrigerators.
These cryogenic test facilities, combined with existing large bore, high field solenoids
in Fermilab's Technical Division, will enable detector and materials testing in an
extended range of temperatures, field strengths, and detector sizes beyond that
which is currently available using existing facilities.   

The detector group at Fermilab is performing research in the area of optical MKIDs,
CMB bolometer arrays, SuperCDMS iZIPs, Skipper CCDs and qubits as axion detectors.
The lab is currently expanding its scope in the realm of sub-eV particle interactions,
both in the area of sensor development as well as systems integration with highly
multiplexed readouts, pushing these precision energy detectors to even lower threshold
and lower noise operation. The scientists and engineers, who are leading the detector R\&D
in these areas, are actively involved in the experiments where these technologies are
being used, such as SPT-3G, CMB, SENSEI and ADMX. This experience, along with existing
R\&D collaborations with quantum computing experts at the University of Chicago and
material scientists at Northwestern University, and partnership with nearby Argonne
National Laboratory, positions Fermilab scientists well to lead the development of
the next generation of detectors.

\section{LANL}

LANL, in common with most of the National Laboratories, has a large number of Ph.D.
scientists and engineers, which give the Laboratory greater scientific breadth and depth
than any university.  The Laboratory can assemble multidisciplinary teams that include
expertise in areas such as quantum information science and quantum metrology, nuclear
physics, particle physics, materials and condensed matter physics, engineering, 
computer science and high-performance computing.  The Laboratory has a long and
distinguished history in the area of quantum information science, including
pioneering work in decoherence, NMR quantum computing, and quantum key distribution
and quantum communication.  Today, LANL's recently-acquired D-Wave 2X adiabatic
quantum computer forms the core of an growing effort to understand what types of
problems will benefit from quantum computing and what performance gains might be possible.

In addition to research centered on the D-Wave machine, LANL supports theoretical
programs in decoherence, quantum algorithms, quantum simulation, experimental
programs in matter wave circuitry for quantum sensing, in quantum communications, and
in atomic magnetometry.   LANL has several research programs in the general area of
high energy physics.  First, LANL is involved in the Short Baseline Neutrino (SBN)
program at Fermilab. SBN will consist of 3 liquid argon (LAr) time-projection
chambers (TPCs) located at different distances from the neutrino source.
The goal of SBN is to prove whether or not short-baseline neutrino oscillations are
occurring, and if so, to measure the oscillation parameters. Short-baseline oscillations
would imply the existence of sterile neutrinos, which are fundamental particles that
do not interact by the weak force.  Second, the High Altitude Water Cherenkov (HAWC)
Gamma Ray Observatory surveys the sky for 100 GeV to $> 100$~TeV $\gamma$-rays with 300 Water
Cherenkov Detector tanks covering 20,000~m$^2$ on Sierra Negra Volcano, Mexico.
HAWC enables an indirect dark matter search from $\gamma$-ray annihilation and decay, the study
of quantum gravity effects on propagation of $\gamma$-rays, and of particle acceleration
in extreme magnetic and gravitational fields.  Since both experiments would benefit
from improved photon detectors, they represent an opportunity for quantum sensing to
contribute to HEP.  Third, LANL has a unique Ultracold Neutron (UCN) facility that is
mainly focusing on fundamental neutron properties such as the neutron lifetime and
decay parameters as a vehicle to examine physics beyond the standard model.
Opportunities exist to examine the quantum properties of UCNs, building on
demonstrations of quantum physics under gravity.  Quantum non-demolition detectors
could revolutionize imaging and tracking technology important to many of these
experiments.  Finally, the team developing state-of-the art atomic magnetometers
is collaborating with some of the Laboratory’s particle physicists to apply this
technology to searches for dark matter.

\section{LBNL}
The role of the national laboratories in quantum sensing are two-fold. In the first place,
the national laboratories develop and maintain facilities with unique capabilities that
enable leading experiments and R\&D, both in-house and in the wider community of facility users,
in particular from universities. Secondly, the labs bring \RandD or small experiments to
larger scales, and provide technical expertise, both in terms of scientists and engineers,
in relevant areas of technical development and blue-sky R\&D.

At LBNL, the Molecular Foundry user facility~\cite{lbnl-foundry} provides microfabrication
equipment and related expertise to enable \RandD relevant for quantum sensing.
The ALS user facility~\cite{lbnl-als} provides unique analytic capabilities and
expertise for investigation of quantum materials and phenomena.
The MicroSystems Lab  at the LBNL Physics Division has developed a model of
industrialization which has been very successful for deployment of large
quantities of fully depleted CCD's in large-scale cosmology experiments (DES, DESI).
This model is now being applied to the production of superconducting devices
for next generation CMB experiments. As a next step the laboratory envisions applying this
model to scaling up quantum sensor devices for future HEP experiments.

\section{LLNL}

Detectors for both dark matter and neutrinos have steadily increased in sensitivity
over the last several decades by orders of magnitude and are beginning to reach limits
set by fundamental quantum mechanics. Examples include near quantum-limited linear
amplifiers for axion dark matter searches and single electron readout of noble liquids
in WIMP/neutrino detectors. At the same time rapid advances in superconducting quantum
technologies, driven in part by the push to develop quantum computing, provide new
opportunities to explore the sensitivity frontier. LLNL, as a cross-disciplinary lab,
has a long history of leveraging national security focused detector \RandD for HEP projects
and vice-versa. This includes being founding members of the ADMX and LUX/LZ dark matter
projects, the LSST telescope and the Prospect, Watchman and nEXo neutrino projects.
LLNL plans to continue this trend by applying infrastructure and expertise developed
for superconducting qubit \RandD towards HEP missions such as dark matter searches.
Additionally ongoing work on cold-atom interferometry~\cite{Dubetsky}, new materials
such as topological insulators and investigations into $^{229}$Th as a potential
material for a nuclear clock could greatly impact HEP by opening up new avenues to
detectors and precision measurements of fundamental constants whose deviations from
the standard model could reveal new physics. 

Pushing the sensitivity frontier involves developing new detectors in close collaboration
with simulations and material modeling groups and exploring new methods of manufacturing.
One recent example is the demonstration of 3D printed superconducting cavities made out
of Ti-6Al-4V (Ti-64), which have surprisingly high kinetic inductance and could be developed
into new quantum microwave amplifiers or photon counters in the THz to gamma-ray
range~\cite{Holland}. Research on novel superconducting cavities could also be
applied to axion searches which require maintaining a high quality factor in the
presence of a large (few Tesla) magnetic field. Additionally microwave single photon
counters applied to axion detectors could dive below the Standard Quantum Limit and
greatly increase search rates by factors of $> 100$ relative to currently deployed
linear amplifiers~\cite{Lamoreaux}. Single photon counting in qubits has been
demonstrated but additional work is needed to mitigate high dark counts and
the challenge of operating near a large magnetic field. Investigating and
mitigating sources of noise such as two-level systems, or in developing new
methods to use photon correlation to suppress noise, is one of the key research
goals of the LLNL quantum coherence device group.

LLNL is also heavily involved in noble liquid detector development.
Examples of recent breakthroughs includes the first demonstration of detection
of sub-keV electron recoils in Argon~\cite{Sangiorgio}.
The challenge of achieving even lower thresholds
involves investigating trapped surface states and ways to mitigate them.
LLNL plans to pursue this \RandD using infrastructure developed over the last decade
with applications to increased sensitivity to WIMP dark matter and neutrino searches
as well as for nonproliferation.

In conclusion quantum sensors are poised to usher in a new landscape for HEP
by pushing beyond current detection limits and LLNL will continue to strive to
leverage national security based investments to open up new discovery space for
HEP missions in dark matter, neutrino physics, photon detection and precision measurement.

\section{NIST}

From early investigations of laser cooling to recent loophole-free tests of Bell
inequalities, NIST has played a central role in the development of many quantum-based
technologies.  This research is motivated fundamentally by the NIST mission of
advancing metrological standards, which often stands at the intersection of pure
science and technology development.  One key in past successes at NIST has been a
close partnership with universities, most notably in the form of the collaborative
institutes. JILA is a decades long
collaboration with the University of Colorado and the Joint Quantum Institute (JQI)
is a more recent partnership with the University of Maryland, both of which are
leading efforts in developing quantum technologies.

The technologies pursued at NIST, including efficient photon detectors,
optical atomic clocks, squeezed light, quantum simulation and quantum computation,
have all had an impact worldwide.  The maturity of these and other quantum
technologies is such that NIST can now envision exploiting quantum correlations
to improve measurement standards.  As this work progresses, NIST will continue
to play an important role not only in developing quantum science in its own
laboratories but also in transferring those developments to other applications
that are, strictly speaking, outside of the scope of NIST's work.
Among these applications there is ample overlap with measurements in
fundamental physics.  Examples include sensitive detectors for the cosmic
microwave background, searches for ultralight dark matter (e.g. axions,
dark photons, dilaton-like dark matter), and measurements of the possible
time variation of fundamental constants.

\section{ORNL}
In recent years, quantum optics has undergone a renaissance in terms of its applicability
to various sensing scenarios.
Two major ingredients of quantum optics, entanglement and quantum noise reduction
(or squeezing)~\cite{slusher_observation_1985,caves_quantum-mechanical_1981}, have found
their way into many sensors that now beat the equivalent state of the art classical sensor.
Advantages of the squeezed light approaches include the ability to reflect squeezed light
off of many different transducers to achieve sensitivity improvements relatively quickly.
Squeezed light-based quantum sensors,
such as SU(1,1) interferometry and direct use of squeezing, probably represent the lowest
hanging fruit in
quantum sensors (for example as in LIGO).
ORNL has produced a bevy of quantum sensors over the past
decade~\cite{Fan, Lawrie, Lukens, Lawrie1, Williams, Williams1, Pooser, Pooser1, Otterstrom}. 
In several cases these represent world records for sensitivity in their respective fields.
In quantum optical sensing specifically, these sensors are the state of the art for
differential detection, with the largest degree of quantum noise reduction demonstrated
in any optical sensor. Two of these efforts, focused on magnetometry~\cite{Otterstrom} and
beam displacement measurements~\cite{Pooser} below the Standard Quantum Limit, are directly
relevant to current axion and dilaton-like dark matter searches.  In addition, ORNL has
expertise in hyper-entanglement~\cite{PhysRevLett.95.260501}, which has been recently
shown to enable sub-shot noise sensing~\cite{PhysRevA.97.010301}.

It is also likely that future quantum sensor networks will be comprised of multiple,
disparate small-scale quantum
devices. This can be achieved by linking individual quantum sensors using photons as
a quantum interconnect to allow quantum information exchange between subsets of sensing
devices. Popular matter-qubit systems, often used as quantum sensors, such as trapped ions,
neutral atoms,
quantum dots, single dopants in silicon can be readily coupled to optical photons.
ORNL recently developed and  demonstrated a universal quantum information networking
protocol for photonic qubits encoded using narrow frequency
bins~\cite{lukens_frequency-encoded_2017, lu_electro-optic_2018}. The protocol
offers linear scaling of optical resources with off-the-shelf telecom components
(as opposed to quadratic scaling of conventional
spatial/polarization-encoding-based protocols) and enables highly parallel entanglement of
multiple frequency modes within a single optical fiber.
Therefore, the photonic frequency degree of freedom is a promising route to
construct an entanglement-swapping-based interconnect for frequency-mismatched
matter qubit sensors.

Superconducting sensors can also be networked via continuous variable (amplitude or phase)
interconnects. These variables are precisely what are entangled in optical fields
when quantum correlations that yield quantum noise reduction are generated. ORNL has
constructed continuous variable quantum networks based on transduction of optical fields
to multiple spatial modes and to electronic nanoscale networks. Based on recent
demonstrations of long range, continuous variable entanglement
networks~\cite{pooser_continuous-variable_2014, yoshikawa_invited_2016, chen_experimental_2014, lawrie_robust_2016}, the laboratory is now poised to construct quantum sensor networks of atomic clocks,
magnetometers, or superconducting detectors  with enhanced sensitivities that approach
the limits of quantum mechanics.

Beyond quantum sensing with optical fields, ORNL also hosts one of the
world's brightest neutron sources at the Spallation Neutron Source (SNS), which can be
used to probe the fundamental properties of myriad materials. The source has also
been used to generate neutrinos as candidate for the hard-to-detect coherent elastic
neutrino-nucleus scattering process~\cite{Akimov}.
The record-breaking detection rate of these
neutrinos at ORNL enabled the first observation of this scattering process~\cite{Akimov}.
The SNS has also enabled a unique neutron EDM experiment~\cite{Golub} in which neutrons are
cooled and stopped via scattering with ${^3}$He. The $^{3}$He also serves as a
magnetometer which detects the signature of the neutron EDM.

\section{PNNL}

PNNL has a number of capabilities that can be brought to bear on quantum information
science challenges over the coming years.  Common to the field are challenges in
material science - both for quantum computing and quantum sensing.
PNNL has a robust program in atomically precise materials including molecular beam
epitaxy of thin films and hetero-structures of complex oxides, development of computational
quantum chemistry tools and a range of metrology tools for sample characterization.
Chemical impurities and other defects limit performance of quantum systems.  
On the computational side, PNNL's NWChem quantum computing code running on HPC platforms
can be used to model individual atom or lattice defects at the quantum level.
That code is currently being revamped for use on heterogeneous exascale computing
(e.g. GPUs) and is planned to be expanded to run using quantum co-processors.
This code base is similar to that required for QCD calculations and the expertise and
experience developing this code for exascale and quantum co-processing may be readily
transferrable to development of QCD calculations on these platforms.
PNNL's experience in ultra-trace detection of both stable and radioactive contaminants
and preparation of exquisitely pure materials for radiometric counting, neutrinoless
double-beta decay and dark matter detection can be employed to address materials
challenges in QIS.  PNNL is a member of the ADMX dark matter and Project~8 neutrino
mass experiments playing a critical role in microwave engineering to detect the very
faint signals that are subsequently amplified using quantum sensing techniques.
PNNL is also acquiring a high-capacity dilution refrigerator suitable for testing
cryogenic dark matter detectors such as those for SuperCDMS, as well as providing
test stand capability for QIS sensing - a capability that currently is capacity-limited
in the complex.

\section{SLAC}
SLAC works to apply quantum sensors to enhance the sensitivity of High-Energy Physics
experiments, especially in the areas of tests of physics beyond the Standard Model
and probing dark matter and dark energy.
One of the earliest applications of quantum sensors to HEP science is
in the identification of dark matter and exploring the unknown, two of the
Particle Physics Projects Prioritization Panel (P5) science
drivers.   SLAC has efforts to apply quantum sensors for precision measurements
of fields below the Standard Quantum Limit (SQL), and to use quantum sensor systems
to improve sensor networks.

The microwave-cavity axion community has already begun to deploy schemes to
evade the SQL, by the use of squeezed-vacuum states.
Covering a larger mass range of dark matter candidates will require developing
quantum sensors exploiting multiple forms of quantum correlations, including
squeezing, entanglement, photon counting, and backaction evasion. SLAC plans
to collaborate with other laboratories and the academic community on this effort,
with an initial focus on quantum sensor technology for use in the
100~Hz - 300~MHz dark-matter mass range. This effort uses frequency
upconverters based on microwave SQUIDs to improve limits on axions and hidden photons. 

The true impact of quantum sensors on HEP science will only be reached when full
quantum systems are implemented. Quantum sensor systems require quantum techniques
not just at the level of individual sensors, but also in their integration, readout,
networking, and data acquisition. Even if the output of the individual elements is
limited by noise from the source (e.g. pixels in an astronomical camera), quantum
networks can sometimes be used to instrument larger, higher performance sensor
arrays with reduced system resources. Quantum networks can play an important role
with superconducting sensors, which can be implemented in HEP experiments to
search for dark matter (e.g. light WIMPs), to study signatures of inflation,
to search for light relics, to constrain neutrino mass with the CMB, and in
future dark energy experiments.

One area of quantum networks that is ripe for use by HEP is in large arrays
of superconducting sensors, including transition-edge sensors (TES). Squeezing
and entanglement can be used to improve the performance of TES arrays, even if
the performance of single pixels is limited by noise in the source. When TES
detectors are read out using microwave SQUID frequency upconverters,
the signal from a TES is encoded only in the phase of the resonator, and not in
the amplitude. Present microwave SQUID TES readout electronics measure both
the phase and amplitude of the signal in the microwave resonator, so they are
limited by the SQL. However, if techniques including
squeezing and multi-resonator entanglement are implemented, the phase of the
signal can be measured below the SQL. Quantum protocols reduce the noise of
the measurement of the TES, which increases the effective dynamic range of
each pixel, reducing the effective bandwidth required by each pixel,
increasing the multiplexing factor, and reducing the power required in
the resonator. SLAC has an effort both to develop these protocols, and the
room-temperature RF electronics required for quantum control of large arrays
of these devices. Quantum sensing will make it possible to make larger,
more powerful arrays of superconducting detectors with less system resources.
 
\chapter{Conclusions}
\label{chap:concl}


\bigskip

The workshop presented a broad and compelling program of ongoing and
future activities in applications-focused quantum information and sensing science.
This work is carried out by a diverse group of researchers, in universities and
labs, and supported by the complete triad of sponsors, i.e. federal agencies,
scientific philanthropy and the commercial sector.  All this attests to a
robust national capability, witnessed by the awarding of the first Nobel
prize explicitly for quantum information science to David Wineland and Serge Haroche in 2012.
The foundations are in place for ambitious initiatives for the broad exploitation
of quantum sensing and information as DOE and NSF have recently embarked upon.
Since the workshop brought practitioners from a broad array of sciences together,
it was an mind-expanding experience for many participants. Many of the techniques
and technologies currently being used in the area of quantum sensing are
unfamiliar to many in the particle physics community.

The panel discussions and interventions by the workshop participants uniformly
supported the assessment that great opportunities exist for ambitious new
initiatives, and provided important thoughts and guidance on the
path forward for OHEP, contained in this report.  Among the conclusions of
the workshop however, were some valuable qualifying or cautionary notes, to
avoid the possibility of misplaced expectations or that such a program would
be oversold early on.  Specifically, it is important to avoid the naïve notion
that quantum information and sensing science represents a panacea for every
sector of high energy physics; it certainly does not.
It will also be
important for OHEP to take stock of what is already being done with these
tools in fundamental physics, and by whom, and what the implications are
of the implementation of these tools in the field, for the experiments where they would be
relevant and useful.

Given the potentially far-reaching implications of a major Quantum
Information and Sensor initiative, both due to the transformative and even
disruptive reach of this technology, and the structural changes that could
ultimately be set in motion, it is crucial that the community embark on such
an initiative with as complete a knowledge of the current landscape and a
recognition of how it may be different from the past.  This landscape may
appear somewhat unfamiliar to the traditional world of high energy physicists,
but is a sign of the times of the rapidly evolving world of this new technology.

{\bf Suggestion 1: Deep Survey} \\
In regard to activities in applied Quantum Information and Science, the workshop
revealed a much more interesting and complex landscape than most imagined.
Before formulating a strategy for the field, it
will be crucial to understand what is happening, who the practitioner are, and from where the derive their support.
Accordingly we propose to consider carrying out a deep survey of the current level
of personnel, investments and activities in this area of science. 
Given the diversity of players, it will be important to survey the
field broadly, including practitioners supported by DOE (including HEP, NP and BES),
the NSF, and the private sector including scientific philanthropies.
The survey should produce a matrix of FTEs by category of institution
(universities, labs, private sector), funding sources, and trends.  Such a survey
should have the capability to assess the collaborative nature of this work, i.e. teams
which bridge different offices within DOE/SC for example, international
collaborators, etc., the size of these efforts and how they are evolving.

This is a complex and difficult job to do meaningfully; it is worth considering
whether a body such as the APS would be better equipped, and could be enlisted
for the task.  It is also worth considering whether the NAS could be retained
to conduct a study of the potential impact and retrospective study of QIS in high
energy physics, for example.

The QIS landscape presented at the workshop may have appeared somewhat unfamiliar
to the traditional world of high energy physicists.
Conversely, the workshop provided several QIS practitioners with a valuable
opportunity to become more familiar
with the principal unanswered questions in high energy physics.
The opportunity to
become more fluent in another science discipline was tremendously valued by the
participants. The multidisciplinary learning opportunities that were provided
for scientists at all levels in their career has been noted as a potent
vehicle to foster connections between the fields.
With the accelerating pace of development in many fields of science and the
difficulty in keeping informed about the successes elsewhere that could have
important benefits to particle physics, there is a clear need for
cross-communication.
The most exciting developments will certainly be those that have yet to be
thought of. We suggest that this could be realized by increased communication
between condensed matter and AMO physicists and their HEP colleagues.

{\bf Suggestion 2: Multidisciplinary Workshops} \\
The high energy physics community typically meets and interacts  at the large high energy physics
conference series, such as the meeting of the APS Division of Particles and Fields (DPF),
the International Conference on High Energy Physics (ICHEP) or the Lepton-Photon
conference. These meetings provide little to no intersection with other fields of
science. For the QIS for HEP initiative to be successful it is critical to have
joint, multidisciplinary workshops to provide expanded opportunities for
interaction between particle physicists and QIS scientists. In addition regular 
remote participation meetings that would
serve as a discussion forum for members of the national labs and universities
from both the QIS and HEP communities interested in developing the potential
of QIS for HEP and HEP for QIS would also help to build a new community and
help incubate new ideas at the QIS--HEP interface.

{\bf Suggestion 3: Establish Early Career Opportunities} \\
At the outset of this initiative, an important signal to the community for the beginning
of an enduring capability-building program will be to attract the attention of the talent just
entering the field.  The DOE, along with the NSF should study and implement as soon
as possible, a system of prestigious (possibly named) fellowships, thesis prizes and
early career awards specifically for Quantum Information Sciences applied to
High Energy Physics. These programs should be inclusive of students, postdocs
and early career faculty and laboratory scientific staff. 

For those who are in high energy physics, the lack of multidisciplinary training
opportunities has been noted as another obstacle to fostering connections between
the fields. Given the pace of development in QIS and the fields it draws upon,
there is a clear need for cross-training. As noted earlier, 
joint workshops, multidisciplinary training opportunities, and
expanded opportunities for interaction between particle physicists and other
scientists is required to enable high energy physics to leverage the advances in QIS.

{\bf Suggestion 4: Training} \\
It is offered for consideration that the office of high energy physics
in conjunction with other science disciplines, organize topical schools in the 
area of quantum sensing, analogous to the well established schools in instrumentation and
computing,
to introduce the fundamentals of quantum sensing especially to younger
physicists.

{\bf Suggestion 5: Invest Now In Areas Where QIS Can Impact HEP} \\
Although it is suggested that a deep survey of quantum sensing is carried out
to develop a long-term
research path for bringing the benefits of quantum sensing to high energy physics,
it is clear
that QIS has already an impact, albeit limited, on HEP research.
A suggested action is that DOE and NSF and others continue to invest in these areas of QIS
to accelerate the advancement of the incorporation of QIS techniques and
technologies into high energy physics.

In regard to what areas of high energy physics will be the early adopters
and beneficiaries of the quantum paradigm, the consensus of the meeting
was that it would be physics where the object of the search or probe was most
naturally described as a field rather than a particle.  A good example
of this would be the class of experiments in the light dark matter world
(e.g. axions, hidden photons) where the signal would result from the
coupling of two harmonic oscillators, i.e. the dark matter field and the
electromagnetic field through the medium of a microwave resonator and/or
a magnetic field.  Indeed the microwave cavity axion community has already
exploited near-quantum-limited detectors in operational experiments,
and begun to deploy schemes evading the Standard Quantum Limit,
by squeezed-vacuum states, and soon, qubit-based single-quantum detection.
Clearly, it would reveal the poverty of our imagination if we claimed
that only light dark matter searches would profit from the quantum paradigm,
but for now, it represents the ``low hanging fruit" that is already being picked.
Another example is the tremendous reach of EDM experiments. Future experiments
would benefit from national laboratory infrastructure ans we approach the PeV
range of sensitivity with expected unity CP phase. Working from the nascent
experimental effort already present in national labs, one could accelerate
the development and field new experiments that have discovery potential in a
range that will not be achieved by any conceivable accelerator. Even in the area
of flavor violation EDMs could outshine other approaches. 
Moving forward, new approaches need to be developed for a successful marriage
of quantum information science and high energy physics. It will require
significant studies to frame the fundamental particle physics questions
using quantum sensing techniques. New technologies might have to be grafted,
optimally correlated states will have to be defined, and how squeezing,
superposition or entanglement can be incorporated in the experiments to reach
the Heisenberg limit demand serious investments both in intellectual as well
as capital resources bringing the right communities together.

{\bf Suggestion 6: HEPAP Subpanel} \\
The formation of a HEPAP subpanel is encouraged to study and explore a possible
program for enduring support of Quantum Sensing in High Energy Physics,
and what the appropriate configuration
of such a program should be.  The subpanel would catalyze a first necessary
discussion between DOE and NSF.

Even where quantum sensors are natural and attractive
solutions to significant advances in sensitivity, it would be naïve to think of
them as ``drop-in retrofits'' for conventional detectors on the back end of an
experiment.  Rather, their implementation will most often demand a
holistic rethinking and redesign of the experiment, a good case study being
the incorporation of SQUIDs and  Josephson Parametric Amplifiers in the
microwave cavity experiment, such as ADMX and HAYSTAC,
which required a major effort to design a
large volume field-free zone, in close proximity to a high field
superconducting magnet, where the fringe field needed to be suppressed by
many orders of magnitude.  Likewise the advent of single-quantum detectors
for the axion experiment will require a parallel, and likely equally
challenging development of superconducting microwave cavities that can
work in the main field to reduce their bandwidth to a minimum.
OHEP thus needs to countenance a program where investments in the quantum
sensing aspects must be matched by investments in R\&D in the more conventional
aspects of the experiments; QIS will not bear fruit in isolation.

The workshop also brought to the fore that high energy physics will most likely
impact quantum information science more in the immediate future than the other
way around. High energy physics has unique capabilities in scaling experiments
and delivering large-scale systems where longevity and uniformity in sensor
response is at a premium. These aspects are not currently well aligned with fundamental
quantum sensor research but if the create the conditions to align them it will provide an opportunity for high energy physics to
make significant contributions to QIS while at the same time having the benefit of  becoming 
more conversant in QIS.  

{\bf Suggestion 7: HEP for QIS} \\
Identify those quantum information science areas where high energy physics
can advance QIS in a
unique way, building on its strengths, for example, in scaling and system
integration.

It was not directly obvious during the workshop where the new quantum
technologies could be implemented in the flagship high energy physics
experiments such as the collider experiments at the LHC or the noble
liquid neutrino and dark matter experiments. Identifying such a large-scale application
would, however, be very intriguing and help advance both fields. 

Finally, what was arguably most impressive at this workshop, was to see how the
AMO community has truly pushed the limits in fundamental physics;
eEDM experiments are setting the strictest limits by far on various
superparticles, and atomic beam interferometry is severely constraining
new scalar fields in the Universe.  It is important to realize
that in these cases the tools themselves are the experiment, and
the AMO community is
carrying this out largely without the participation or support of
high energy physicists or their funding.  In launching a Quantum Sensing
initiative, and the solicitations that will eventually derive from it,
OHEP will need to come to terms with
important sociological questions, specifically, who is ``family'' and
who should be eligible for funding? Discovery potential and resources
should have a strong correlation for the benefit of the science community
and society. 

An overarching issue that could overshadow the tremendous opportunity that the
adaptation of quantum sensing promises for particle physics is that there are
still many challenges to initiating, identifying resources to support, and
carrying out multidisciplinary research. Sufficient mechanisms do not currently
exist to enable particle physicists to work collaboratively with other fields
on projects of common benefit.
It is these areas that could define, or at least have a significant impact,
not only the future of particle physics but, increasingly, the future of other
connected fields and segments of industry.

{\bf Suggestion 8: Develop Mechanisms To Support Interdisciplinary Research Teams} \\
Funding for fundamental research unfortunately remains very stove-piped. For a quantum
sensing initiative, which is inherently multidisciplinary, to be successful mechanisms
need to be developed to fund multi-PI research teams for specific term-duration projects,
short-term personnel exchanges, and opportunities for senior researchers to spend a
well-defined percentage of their time on ideas outside the mainstream of
particle physics and its traditional
technologies R\&D. 

Scientific philanthropy, previously associated in the physical
sciences mostly with large telescopes, has played an
increasingly strong and effective role in Particle Astrophysics.  Foundations such as
the Simons, Heising-Simons, the Keck, and the Gordon and Betty Moore foundation
have made increasingly large and strategic investments in specific subfields of
particle astrophysics and cosmology, notably axion dark matter, and cosmic microwave
background radiation studies.
These investments have been guided by thorough and professional scientific advisory
and review processes rivaling those of the federal agencies.  The scale of funding
awarded to some experiments or observatories in some cases have been of the same order
as their federal counterparts, and the partnerships with the NSF have been extremely
fruitful and congenial.  To our knowledge, there has been much less of a tradition of partnership 
with DOE, but this should change.  The Heising-Simons Foundation convened a workshop
at NSF a few years ago reviewing the history of public-private partnerships, which
on the main have been very effective.

{\bf Suggestion 9: Partnerships} \\ 
We encourage DOE, NSF and the major science supporting philanthropic foundations to
begin to discuss how to improve coordination to realize the potential
of science gains for
both QIS and HEP by bringing the fields together.

It is clear, as we believe the deep survey proposed above will document,
that BES has been a strong player in this field, operating largely independently
of the high energy physics community; DOE NP was also in evidence at the workshop.
Beyond DOE, NIST has played a leadership role, perhaps the leadership role in this
field and needs to be engaged.  Also offices and missions
within NASA have invested in QIS technology for early universe cosmology,
and common cause should be made with them. 
The R\&D arms of the services (e.g. ARO, ONR, AFOSR) have all made investments in
this area in recent years.  It is interesting to note that behind NSF (231),
and DOE (115),
ONR comes in third with 64 Nobel Prizes associated with research they have supported,
including several pioneers in the Quantum Information realm, such as
Ted H\"ansch, Jan Hall, Carl Weiman, Eric Cornell, Wolfgang Ketterle, Daniel Tsui,
Horst St\"ormer, William Phillips, Hans Dehmelt, Robert Schreiffer, Leon Cooper,
and Norman Ramsey.
Joining the defense and service labs, IARPA and its predecessors also have a long history of
supporting QIS.  More recently, tech giants such as Google, have launched major
initiatives to complement government investments. 


To conclude, there is great promise in the use of quantum sensing for HEP.
The challenges faced by this emerging interdisciplinary
science may lead to a productive realignment and redefinition of parts 
of both scientific disciplines. 
We confidently predict that a targeted initiative by DOE OHEP will act
as a spur to discover
other areas of high energy physics where quantum information and sensing have the potential to
dramatically reshape a significant fraction high energy physics in the fullness of time.

\bibliography{qs}

\end{document}